\documentclass[ag]{copernicus}

\usepackage{color}

\begin{document}

\title{\textsf{A model of so-called `Zebra' emissions in type IV radio bursts}}

\author[1,2]{\textsf{R. A. Treumann}\thanks{Visiting the International Space Science Institute, Bern, Switzerland}}
\author[3]{\textsf{R. Nakamura}}
\author[3]{\textsf{W. Baumjohann}}

\affil[1]{Department of Geophysics and Environmental Sciences, Munich University, Munich, Germany}
\affil[2]{Department of Physics and Astronomy, Dartmouth College, Hanover NH 03755, USA}
\affil[3]{Space Research Institute, Austrian Academy of Sciences, Graz, Austria}

\runningtitle{Zebra emission theory}

\runningauthor{R. A. Treumann, R. Nakamura, and W. Baumjohann}

\correspondence{R. A.Treumann\\ (rudolf.treumann@geophysik.uni-muenchen.de)}

\received{ }
\revised{ }
\accepted{ }
\published{ }


\firstpage{1}

\maketitle

\begin{abstract}
A simple mechanims for the generation of electromagnetic Zebra pattern emission is proposed. The mechanism is based on the generation of an ion-ring distribution in a magnetic mirror geometry in the presence of a properly directed field-aligned electric potential field. The ion-cyclotron maser then generates a  number of electromagnetic ion-cyclotron harmonics which modulate the electron maser emission. The mechanism is capable of switching the emission on and off or amplifying it quasi-periodically which is a main feature of the observations. 

\end{abstract}

\section{\bf\textsf{Introduction}}
Substantial effort has been invested into the explanation of an apparently miniscule effect in radio emission from some objects, in particular from the Sun during type IV radio bursts. These radio bursts are caused in the aftermath of solar flares by electron-cyclotron maser emission from the anisotropic magnetically trapped flare-electron population. 

`Zebra' emissions are spectral fine structures which appear as narrow-band multi-harmonic modulations (emission/absorption structures) in the type IV radiation \citep[for a recent review cf., e.g.,][]{chernov2010}. Their quasi-harmonic frequency spacing has sometimes been suspected to be related to the ion cyclotron frequency, though in the past a number of mechanism have been proposed which search for explaining these fine structures differently \citep[as for an example see, e.g.,][]{labelle2003} as representations of inherent properties of the electron radiation or other processes like wave-wave interactions. 

Here we exploit the very simple idea that these emissions could indeed be a consequence of the presence of ion-cyclotron harmonics which are genuinely caused by an high energy ion distribution which propagates on the radiating electron background, itself radiating ion cyclotron waves which propagate on the R-X mode branch and, because of their frequency $\omega\sim \ell\omega_{ci}\ll\omega_{ce}$ (with harmonic number $\ell=1,2,3\dots$) being much less than the electron cyclotron frequency $\omega_{ce}$, cannot leave the plasma. Interaction with the type IV electron background under certain restrictive conditions then modulates the electron maser and causes the observed `Zebra' fine structure. 

Justification of such a model (as is shown in Figure \ref{fig-zebra-model}) can be given by reference to the flare reconnection model, according to which flares are caused by magnetic reconnection in the lower corona at field strengths of several Gauss $\lesssim B\lesssim$ few 100 Gauss. In this process localised magnetic field-aligned electric potential drops develop which accelerate and heat the flare electrons and accelerate ions until they evolve into high energy essentially cold conical phase space distributions, a process well-knowm from auroral physics. Under type IV conditions the ions become relativistic. Their cold conical distributions excite ion-cyclotron harmonic waves via the ion-cyclotron maser instability.


\section{\textsf{Ion-cyclotron maser instability}}
The cyclotron maser is a relativistic instability that involves the presence of at least weakly relativistic particles. It had been originally proposed for electrons only, first in the construction of free-electron lasers \citep{motz1951,madey1971} and masers before being recognised for plasmas \citep[by][in its non-relativistic non-efficient version]{twiss1958} and, much later \citep[in its relativistic version by][]{wulee1979}, as a very efficient radiation mechanism in space plasmas as well \citep[for a review cf.][]{treumann2006}. Subsequently it has been identified as the main radio wave emission process in the aurora of Earth and other planets, in solar radio bursts, particularly type IV bursts, and under astrophysical conditions. 

The electron cyclotron maser as the main emission mechanism in all these objects is well established by now. Below we will make use of it. However, as for the Zebra fine structure which interests us here, the electron cyclotron maser is just the carrier of information.  Adopting the notion that the Zebra pattern maps the ion cyclotron frequency into the emitted electron maser radiation, we propose that an exactly similar mechanism works as well for the ions causing an ion-cyclotron maser instability. Such an instability has, in fact, been proposed first by \citet{hoshino1991} and was refined by \citet{amato2006} to include a cold relativistic gyrating ion beam. In a different non-relativistic version it was used in space plasma \citep[cf., e.g.,][]{chaston2002} where it should serve to excite harmonics of low frequency electromagnetic ion cyclotron waves in the auroral plasma. 


\begin{figure}[t!]
\centerline{{\includegraphics[width=0.5\textwidth,clip=]{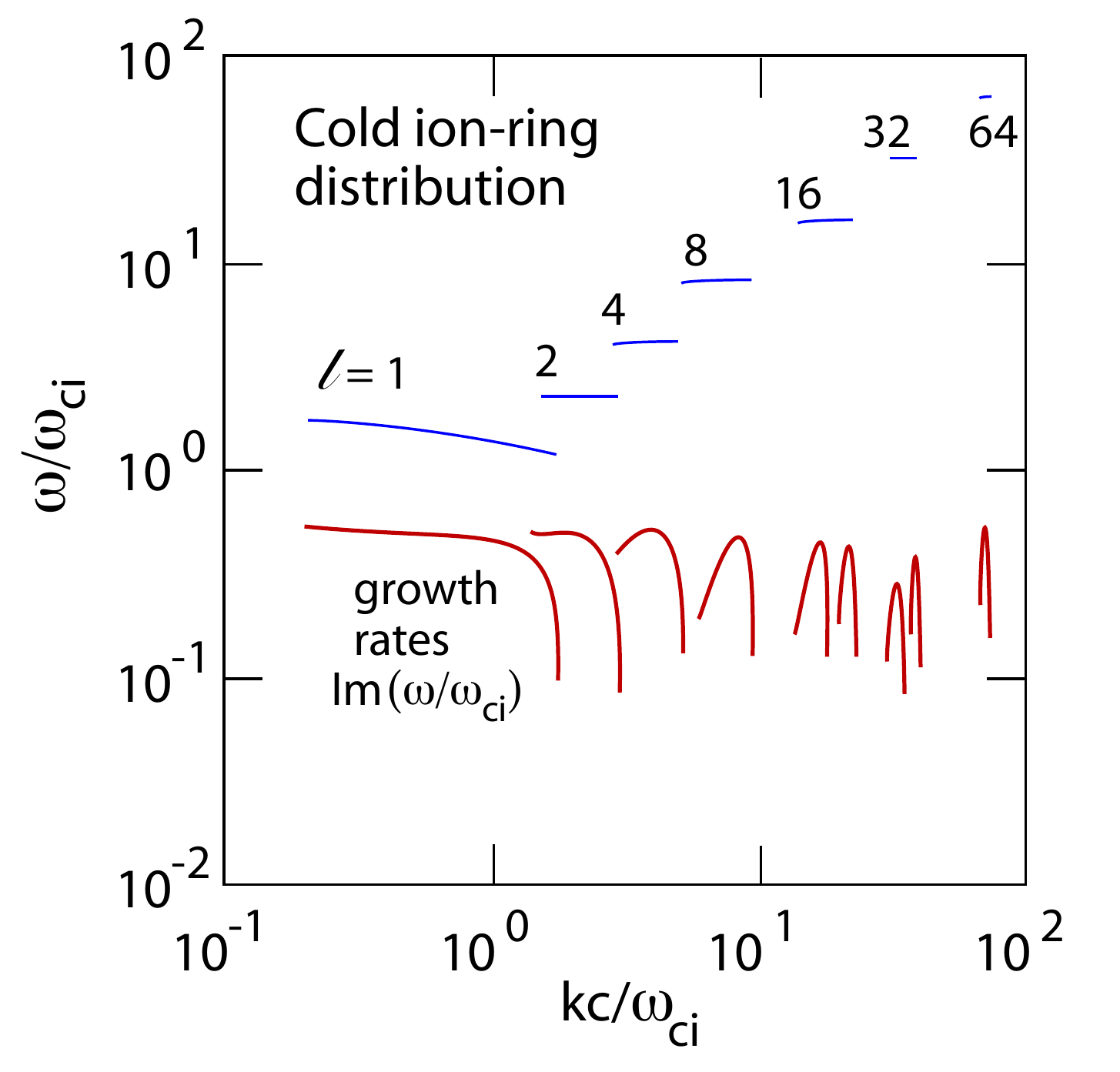}
}}
\caption[ ]
{\footnotesize Ion-cyclotron maser dispersion relation. Harmonic frequencies (blue) and growth rates (red) excited by a cold relativistic proton ring distribution in an electron-proton plasma with ring-Lorentz factor $\gamma_{ib}<10, N_{ib}/N\sim 10^{-2}$ \citep[re-scaled data taken from][]{amato2006}. The ring-ion maser generates a large number of ion cyclotron harmonics $\ell$,  all of nearly same maximum growth rate.}\label{ionmaser-fig1}
\vspace{-0.3cm}
\end{figure}

For efficient generation of ion cyclotron harmonic waves under type IV burst radiation one requires the presence of a relativistic gyrating ion distribution with anisotropic velocity and positive perpendicular phase space gradient. For simplicity, the ion distribution is taken to be cold and anisotropic 
\begin{equation}
F_{ib}(u_\perp,u_\|)=\frac{1}{2\pi U_{ib}}\delta(u_\perp-U_{ib})\delta(u_\|)
\end{equation}
where $u$ is the normalised 4-velocity [$u=\gamma\beta$ with $\gamma=(1-\beta^2)^{-\frac{1}{2}}=(1+u^2)^\frac{1}{2}$ the Lorentz factor], and $U_{ib}$ is the initial value of $u$ for the relativistic ions, and subscripts $\perp,\|$ refer to the direction of the magnetic field. For electromagnetic waves with perpendicular wave vector ${\bf k}=k{\hat x}$ the dispersion relation becomes
\begin{equation}
n^2=\epsilon_{yy}-\epsilon_{xy}\epsilon_{yx}/\epsilon_{xx}, \quad \epsilon_{xy}=-\epsilon_{yx},\quad n^2\equiv k^2c^2/\omega^2
\end{equation}
The explicit form of the dielectric tensor components is standard and has been given, e.g., by \citet{amato2006} whose results we will make use of here and who use a pair plasma as background. In order to exploit their results we have to re-scale their parameters to conditions of an electron-ion background. These parameters are the ratios $\sigma=B^2/\mu_0mNc^2\gamma$ of magnetic to kinetic energy densities for the different species (in their case, the ion beam and background electron and positron pairs, index $\pm$, for which they use numerically $\sigma_{ib}=10^{-2}, \sigma_\pm=2$). We replace the positrons with ions and use the quasi-neutrality condition $N=N_e=N_i+N_{ib}$, where the subscript $ib$ denotes the relativistic ring ions. This yield for the background ion Lorentz factor $\gamma_i=\mu\gamma_e/(1-N_{ib}/N)\approx\mu\gamma_e$. For small beam densities $N_{ib}\ll N$, with $\mu=m_e/m_i$, the background ions are non-relativistic, and only the radiating electrons will be considered relativistic. Moreover, we have $\gamma_{ib}/\gamma_e\sim200\mu (N/N_{ib})$. Since the electron background is assumed mildly relativistic we will have $\gamma_{ib}\sim 0.1 (N/N_{ib})\gamma_e\lesssim 10$ for applying the results of \citet{amato2006}. With these quite reasonable numbers we can directly refer to Figure 13($a$) of \citet{amato2006} the re-scaled version of which which we reproduce here as Figure \ref{ionmaser-fig1}. 

One observes that a very large number of electromagnetic ion cyclotron harmonics is generated which propagate on the R-X mode branch.  These waves have all roughly the same growth rate and are of narrow bandwidth. These harmonics are practically all confined to the plasma even in an underdense plasma under the condition $\omega_e^2/\omega_{ce}^2<2T_e/m_ec^2<1$ \citep{winglee1983} of the electron-cyclotron maser instability as long as their frequency is below the upper hybrid frequency $\omega_{uh}\gtrsim\omega_{ce}$ in this case.  The perpendicular electric fields $E_\perp(t)=\sum_\ell E_\ell\exp(-i\ell\omega_{ci}t)$ of these ion cyclotron harmonics generate drift motions in the magnetically trapped  anisotropic electron background plasma that is responsible for the type IV radiation. These electron drifts modulate the emitted electron cyclotron maser radiation, adding factors $2\pi\delta(\omega\pm\ell\omega_{ci})$ to the emission spectrum. 
\begin{figure}[t!]
\centerline{{\includegraphics[width=0.5\textwidth,clip=]{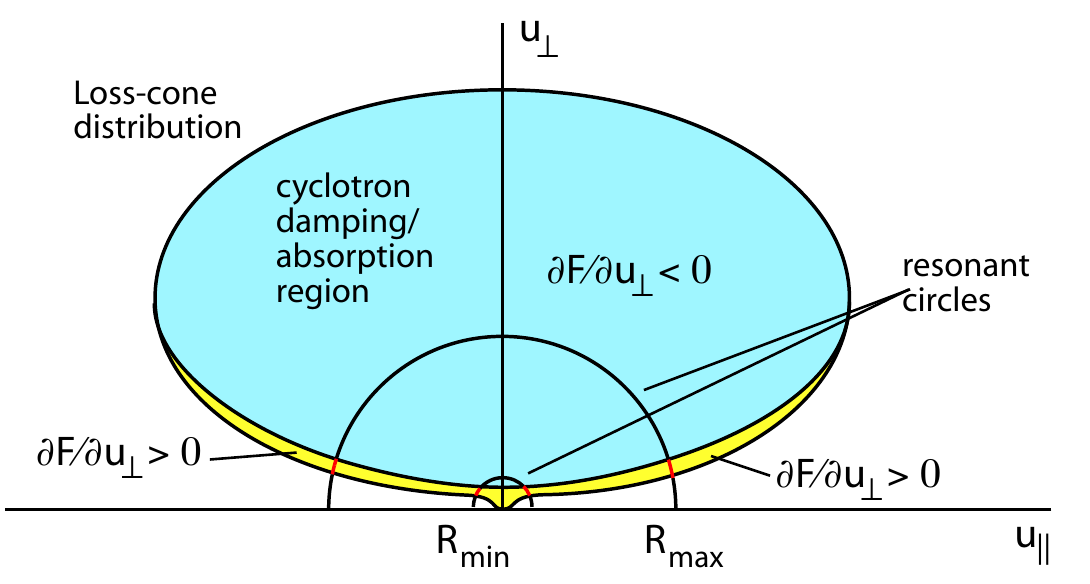}
}}
\caption[ ]
{\footnotesize Electron loss-cone (Dory-Guest-Harris) distribution in the normalised 4-velocity plane. Shown are two resonance circles for maximum and minimum radii, corresponding to minimum and maximum unstable electron cyclotron maser emissions in perpendicular propagation of R-X- modes. The section of the resonance circles contributing to emission are shown in red. Emission is provided as long as the phase space gradient over the positive (yellow)
  part is steep enough to compensate for the larger part of the circles traversing the blue negative gradient region of the distribution. Note that the blue region between the circles of maximum and minimum resonances $R_{\max}, R_{min}$ absorbs electron-cyclotron waves by damping them. This heats the electrons in this domain and flattens the distribution in this region of phase space.}\label{fig-dgh-distr}
\vspace{-0.3cm}
\end{figure}

\section{\textsf{Electron-cyclotron maser radiation}}
We briefly indicate what kind of variation in the spectrum of the electron cyclotron emission is to be expected in this case. The model of type IV radiation is based on the assumption of a magnetically trapped electron distribution. Such distributions are of loss-cone type which can, for instance, be modelled as (relativistic) Dory-Guest-Harris distributions
\begin{equation}\label{DGH}
F_e(u_\perp,u_\|)=\frac{1}{\pi^\frac{3}{2}v_e^3s!}\left(\!\frac{u_\perp^2}{v_e^2}\!\right)^{\!\!\!\!\!s}\!\exp\left[-\frac{u_\perp^2+u_\|^2}{v_e^2}\right], \quad s \in{\textsf{R}}
\end{equation}
where $v_e^2$ is the normalised thermal spread, and the factor $u_\perp^{2s}\propto\sin^{2s}\theta$, with $\theta$ the angle between 4-velocity and magnetic field direction, accounts for the loss cone. Note that $s$ can be any real number while being most conveniently chosen as $s\in\textsf{N}$.

With the above distribution function the instability has been exhaustively discussed by \citet{pritchett1986} for different propagation angles finding that largest growth is obtained for strictly perpendicular propagation, $k_\|=0$. In this case the resonance condition for resonant excitation of waves becomes $u_\perp^2+u_\|^2 -2(1-\nu_{ce})=0$, where $\nu_{ce}=\omega/\omega_{ce}$ is the non-relativistic ratio of wave frequency to electron cyclotron frequency. Clearly, resonance is possible only for $\nu_{ce}\lesssim 1$ and, as usual, the resonance line is a circle in the normalised 4-velocity plane $(u_\perp,u_\|)$ of radius 
\begin{equation}
R_\mathrm{res}=\sqrt{2(1-\nu_{ce})}
\end{equation}
located between the minimum and maximum radii $R_{min},R_{max}$ of resonance (see Figure \ref{fig-dgh-distr}) which correspond to maximum and minimum resonant frequencies, respectively. 

The relativistic cold plasma dispersion relation of the R-X mode,  which is the real part of
\begin{equation}
n^2-1-(2-n_\perp^2)A=0, \quad\mathrm{with}\quad n^2=k^2c^2/\omega^2
\end{equation}
allows for a range of such resonant frequencies below the non-relativistic $\omega_{ce}$ which is quite narrow \citep[cf., e.g.,][]{pritchett1986}, depending on the plasma parameters. The maximum resonance frequency (minimum radius) is very close to $\nu_{ce}=1$. On the other hand, the minimum resonance frequency can be estimated from the cold X-mode dispersion relation yielding 
\begin{equation}
\nu_{ce, min}\approx 1- \frac{\gamma_e^2}{2}\left(\frac{\omega_e}{\omega_{ce}}\right)^2
\end{equation}

The function $A$ contains the plasma response. Taking $k_\|=0, |1-\nu_{ce}|\ll 1$, one has to first order $\mathrm{Re}(A)\ll 1$. This expression is to be used in the calculation of the growth rate 
\begin{equation}
\mathrm{Im}(\nu_{ce}) \simeq-{\scriptstyle\frac{1}{2}}\mathrm{Im}(A)
\end{equation}
Maser radiation will be produced if $\mathrm{Im}(A)<0$ is negative (corresponding to `negative absorption' of the electromagnetic R-X mode). The resonant imaginary part of $A$ is given by 
\begin{eqnarray}\label{Im-A}
\mathrm{Im}(A)=&-&\frac{\pi^2}{2\nu_{ce}}\frac{\omega_e^2}{\omega_{ce}^2}\int\limits_{-\infty}^\infty\int\limits_{0}^\infty{\rm d}u_\|u_\perp{\rm d}u_\perp^{2}\frac{\partial F_e}{\partial u_\perp}\times\nonumber \\
&\times&\delta(u_\|^2+u_\perp^2-R_\mathit{res}^2)
\end{eqnarray}

In dealing with the loss cone distribution Figure \ref{fig-dgh-distr} one realises that the modulation of the particle distribution caused by the presence of the ion cyclotron waves just affects the low perpendicular energy electrons at the loss cone boundary. This is the only region which has positive perpendicular gradient and contributes to wave growth over the narrow part of the boundary crossed by the resonant circle. The longer part inside the distribution sees a weak negative gradient and causes wave absorption which for instability is over-compensated by the steep gradient at the loss-cone boundary.   
\begin{figure}[t!]
\centerline{{\includegraphics[width=0.5\textwidth,clip=]{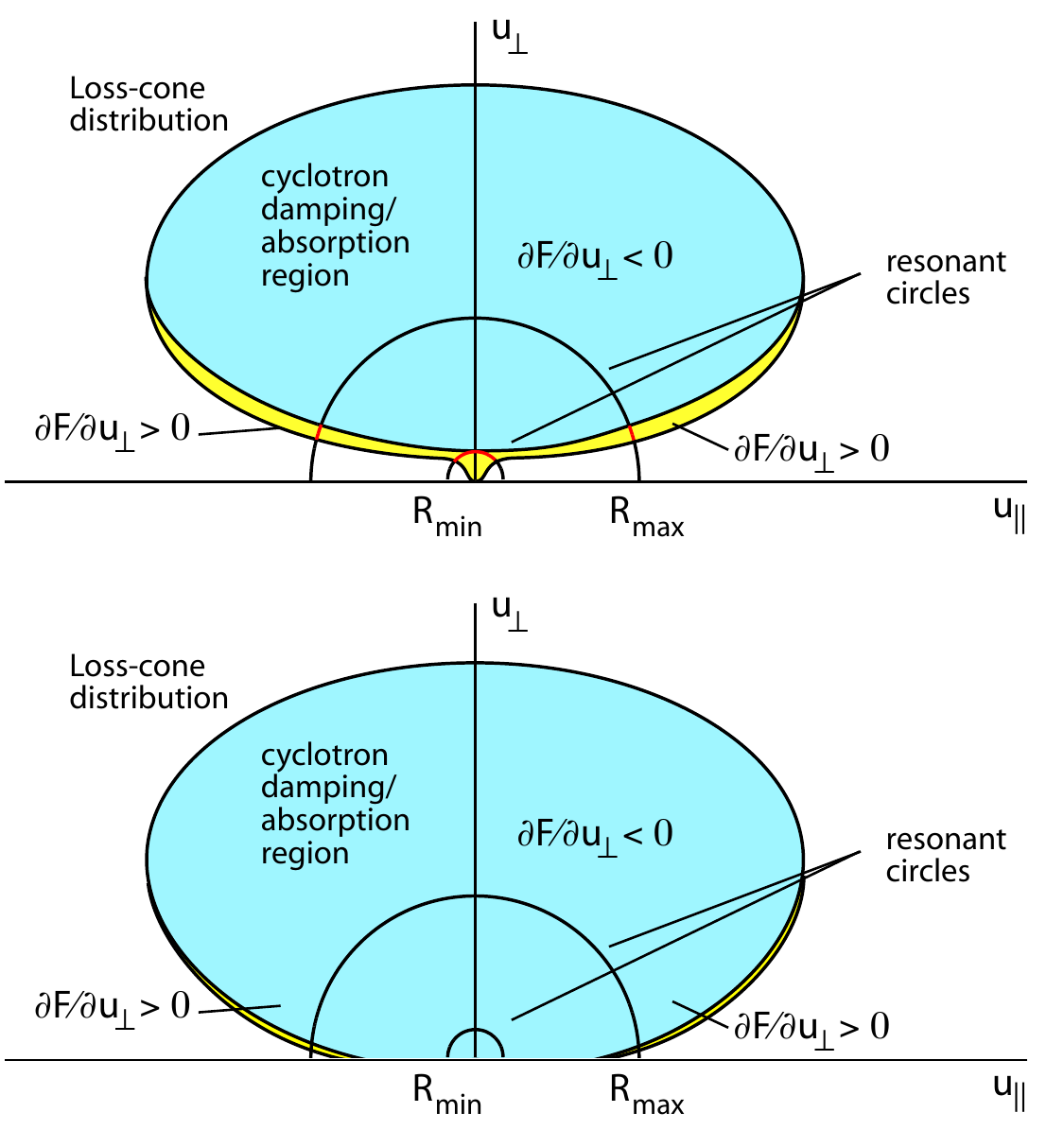}
}}
\caption[ ]
{\footnotesize Same as in Figure \ref{fig-dgh-distr} except for the modulation of the loss cone included. \textit{Top}: The modulation due to the electric field drift in the transverse electric field of the ion-cyclotron mode shifts the perpendicular velocity up thereby opening the loss cone and pushing its gradient to higher $u_\perp$. This steepens the gradient and reduces the damping part of the resonance circle. At minimum radius (largest emission frequency $\nu_{ec}$) the resonance circle is fully in the positive gradient. During this phase emission is amplified. \textit{Bottom}: During this phase the drift reduces the perpendicular velocity filling the loss cone and switching the maser off. }\label{fig-max-min}
\vspace{-0.3cm}
\end{figure}

Inspection of this figure suggests, that the modulation of the loss-cone boundary will heavily affect the emission conditions. In presence of ion-cyclotron waves the low perpendicular energy electrons at the loss-cone boundary will start oscillating up and down in $u_\perp$ with the imposed frequency of the ion cyclotron wave electric field, when performing an ${\bf E}_{\perp}\times{\bf B}$-drift. This drift displaces the loss-cone boundary either up to higher $u_\perp$ or down. In the latter case the loss-cone is partially filled, and the maser may switch off, causing zero emission intensity at the respective frequency. In the former case the loss-cone widens, faking a larger loss-cone exponent $s$ and bringing the minimum-radius resonance almost completely into the positive gradient region which should amplify the emission at frequency very close to $\nu_{ce}\sim 1$.  These two cases are shown schematically in Figure \ref{fig-max-min}.

Including this modulation effect into the calculation of the electron cyclotron maser instability from scratch on is quite involved. We can however boldly construct a simple model which possesses some of the relevant features of the modulation. 

Realising that the two time scales of the ion cyclotron and electron instabilities are quite far apart, we consider the modulation as stationary on the time scale of the electron cyclotron maser emission. Then, the drift effect will result in a modulation of the loss cone as has been described above and can be included by replacing $u_\perp \to u_\perp'=u_\perp+\Delta$ in the loss-cone factor of the electron distribution function Eq.\,(\ref{DGH}), where $\Delta=\gamma_eV_E(t)/c$ is the contribution of the normalised drift in the ion-cyclotron wave electric field.

Modulation can, in addition, also be included as a variation of the loss-cone exponent $s$ replacing it with a variable exponent $s\to a(s,t)>0$ and $t$ the slow time scale, allowing for a direct response of the loss cone to the modulation. Since $a$ is not anymore an integer number, this latter step requires that in the normalisation of the DGH distribution Eq.\,(\ref{DGH}) the factor $s!$ is replaced by the Gamma-function $\Gamma(a+1)$. Performing the derivative yields
\begin{equation}
\frac{\partial F_e}{\partial u_\perp}= \frac{2}{u_\perp'}\left(a-\frac{u_\perp u_\perp'}{v_e^2}\right)F_e
\end{equation}
Inserting into Eq.\,(\ref{Im-A}), performing the integration with respect to $u_\perp^2$, rearranging and neglecting terms higher than first order in $\Delta$ reduces to
\begin{eqnarray}
\mathrm{Im}(\!\!\!\!&A&\!\!\!\!)\approx -\frac{\sqrt{\pi}R_\mathit{res}}{\Gamma(a+1)} \frac{\omega_e^2}{\nu_{ce}\omega_{ce}^2}\frac{\mathrm{e}^{-R_\mathit{res}^2}}{v_e^2}\int\limits_0^1\frac{x^a\mathrm{d}x}{\sqrt{1-x}}\left[\left(\frac{a}{R_\mathit{res}^3}-x \right.\right. \nonumber\\
&+&\left.\left.\!\!\!\!\frac{\Delta'}{R_\mathit{res}}x^\frac{1}{2}\right)
+\Delta'(2a+1)\left(\frac{a}{R_\mathit{res}}x^{-\frac{1}{2}}-x^{+\frac{1}{2}}\right)\right]
\end{eqnarray}
where $\Delta'=\sqrt{1+R_\mathit{res}^2}V_E(t)/c$. All quantities are normalised as indicated earlier. (In order to arrive at the above expression we have taken $R_\mathit{res}$ out and changed the fake integration variable.) The integral can be evaluated making use of $\int^1_0\mathrm{d}x^a/\sqrt{1-x}=\textsf{B}(a+1,\frac{1}{2})$, where $\textsf{B}(x,y)=\Gamma(x)\Gamma(y)/\Gamma(x+y)$ is the beta-function and $\Gamma(\frac{1}{2})=\sqrt{\pi}$. This yields
\begin{equation}\label{growthrate}
\mathrm{Im}(\!A)\approx -C\left(Q_1+\Delta'Q_2\right)
\end{equation}
where $C\equiv(\pi R_\mathit{res}\omega_e^2)\mathrm{e}^{-R^2_\mathit{res}}/(\nu_{ce}\omega_{ce}^2v_e^2) >0$. The term $Q_1$ depends solely on the loss-cone exponent $a$, while $Q_2$ depends on $a$ and accounts for the variation in the perpendicular velocity. These functions are
\begin{eqnarray*}
Q_1&=&\frac{1}{(a+\frac{1}{2})\Gamma(a+\frac{1}{2})}\left[\frac{a}{R_\mathit{res}^3}-\frac{a+1}{a+\frac{3}{2}}\right], \\
Q_2&=&\frac{a(2a+1)\Gamma(a+\frac{1}{2})}{(a+1)\Gamma^2(a+1)}\left[\frac{1}{R_\mathit{res}}-a\right]
\end{eqnarray*}
Observing that always $a\geq0$ we first neglect $Q_1$. Then, 
$Q_2$ [and Im($A$)] changes sign at $a=R_\mathit{res}^{-1}=[2(1-\nu_{ce})]^{-\frac{1}{2}}$. If $a=s$ is an integer, this means that at sufficiently large loss cones and for $\Delta'>0$ the term with $Q_2$ damps the resonance and could theoretically cause absorption. On the other hand, since $|\nu_{ce}-1|\ll1$ this requires quite large $a$. We reasonably assume that $a<R_\mathit{res}^{-1}$ and thus $Q_2>0$. However, the sign of this term also depends on $\Delta'\propto V_E\propto \sin(\ell\omega_{ci}t)$ which oscillates with harmonic ion cyclotron frequency $\ell\omega_{ci}$. As we have proposed, this causes a sinusoidal modulation of the term including $Q_2$ and leads to modulation of the electron maser emission at the resonance frequency. At small $a$, however, we have always $Q_2<Q_1$.  In order for the second term in Im$(A)$ to dominate one requires that with $\beta_E=|V_E^{max}/c|$
\begin{equation}
(2aR_\mathit{res}^2\gamma_e\beta_E)^{-1}<\frac{\Delta'}{\gamma_e\beta_E} \approx |\sin(\ell\omega_{ci}t)|<1
\end{equation}
where, for simplicity, we assumed just one sinusoidal ion-cyclotron harmonic belonging to the entire spectrum of ion waves that are generated by the cold ion conic distribution. The argument of $\sin^{-1}$ must be smaller than one. Hence this condition, combined with the condition on $a$ 
\begin{equation}
R_\mathit{res}^{-1}>a\gg(2\gamma_e\beta_E)^{-1}
\end{equation}
and with $\gamma_e\gtrsim 1$ implies that
\begin{equation}
R_\mathit{res}<2\beta_E \quad \mathrm{or}\quad 1-\nu_{ce}\ll 2\beta_E^2
\end{equation}
These conditions can be satisfied only for sufficiently steep loss cone gradients $a$ and large drifts caused by the ion-cyclotron harmonics. If this is the case, the growth rate Eq. (\ref{growthrate}) of the electron-cyclotron maser instability is periodically modulated during the phases of the ion-cyclotron harmonics. It is amplified during positive phases $0<\omega_{ci}t<\pi$ and switched off during negative phases $\pi<\omega_{ci}t<2\pi$. The frequency of this modulation of the electron-cyclotron maser radiation at the local resonance is proportional to the ion cyclotron frequency $\omega_{ci}$.
\begin{figure}[t!]
\centerline{{\includegraphics[width=0.5\textwidth,clip=]{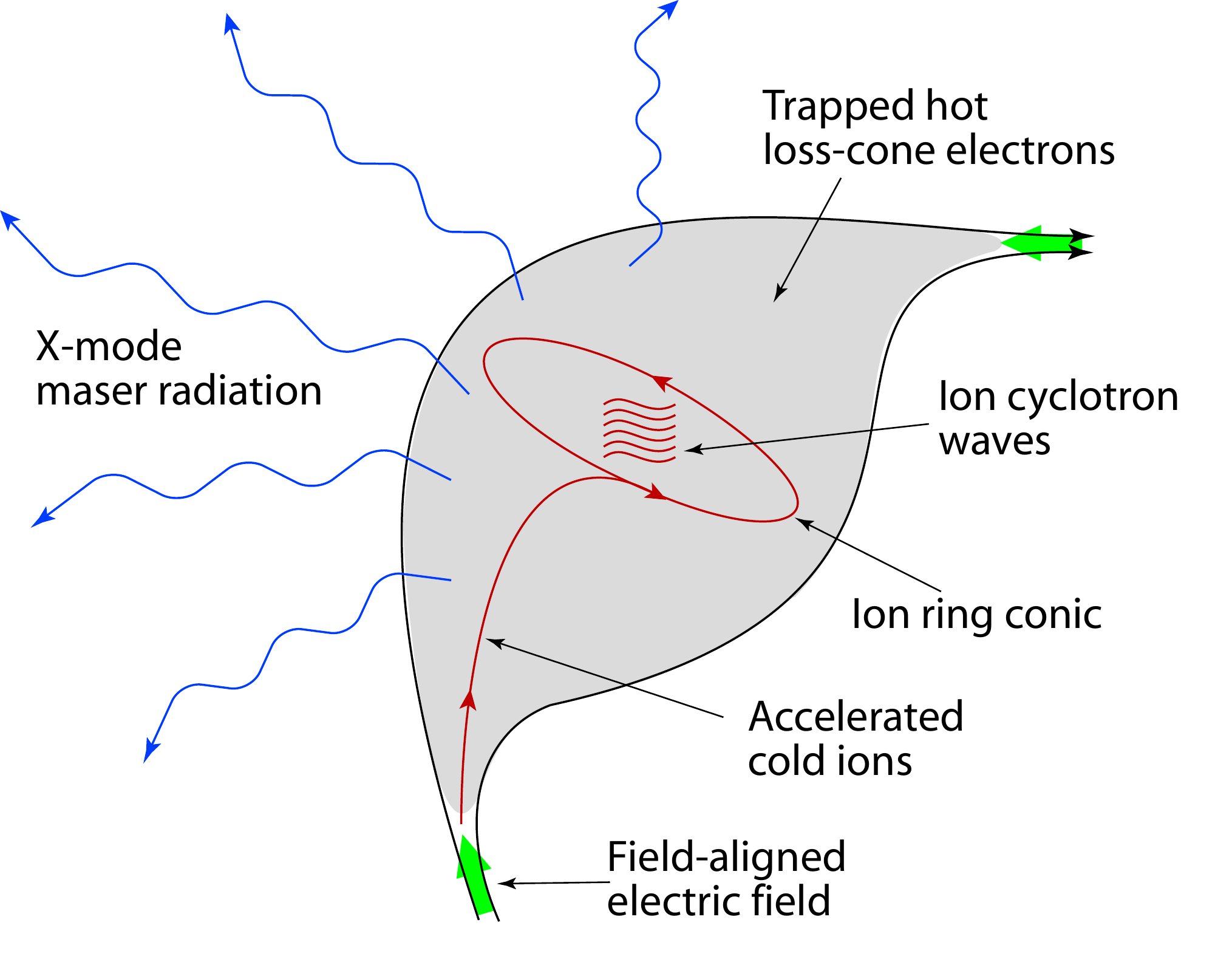}
}}
\caption[ ]
{\footnotesize The model of a post-flare magnetically trapped electron distribution which is the source of type IV electron cyclotron maser radiation. In this stage the weakly relativistic hot electrons have evolved into a loss-cone distribution of Dory-Guest-Harris type with completely empty sharply edged loss cone assuming that whistler activity has ceased (simply because $\omega_{pe}/\omega_{ce}\ll 1$) and quasilinear diffusion into the loss cone is terminated. The R-X mode emission is caused on the section of the resonance circle in phase space that cuts through the steep perpendicular phase space gradient of the loss cone and is mostly perpendicular to the magnetic field at frequency below but very close to the local electron cyclotron frequency. Upward directed strong electric fields at the site of the magnetic mirror accelerate cold ions into an ion-conic distribution which propagates up into the configuration and generates large numbers of ion cyclotron harmonics. The local modulation of the electron loss cone distribution by the ion wave spectrum modulates the electron cyclotron maser and causes the genuine ion-harmonic spaced Zebra emissions.}\label{fig-zebra-model}
\vspace{-0.3cm}
\end{figure}

Though the modulation is small in the growth rate, it modulates the exponential amplification factor of the maser emission which causes a much stronger modulation of the emitted radiation intensity. However, it requires a sufficiently relativistic trapped electron component with steep loss cone gradient in addition to the presence of a cold relativistic ion ring passing across the electron background.

In order to directly see the modulation, we calculate the power emitted by the electron-cyclotron maser at resonant frequency $\nu_{ce}$. Poynting's theorem yields
\begin{equation}
{\cal P}(\nu_{ce},\tau_{esc})=\frac{1}{\mu_0c}|E_\perp(\nu_{ce})|^2_0\mathrm{e}^{[2C(Q_1+\Delta'Q_2)\tau_{esc}]}
\end{equation}
where we have assumed that $\omega_{ce}\nu_{ce}/k\simeq c$. 

We are interested only in the modulation of the power. We therefore do not calculate the power at saturation which requires the precise knowledge of the saturation process. Since in the collisionless case the radiation can presumably freely escape this requires knowledge of the competition between cyclotron maser emission and cyclotron re-absorption along the ray path. This is not impossible to estimate but lies outside the scope of our investigation. So we stay with the linear state, assuming that the power is limited just by escape from the volume, i.e. by the time $\tau_{esc}\sim L/c$, where $L$ is of the order of the linear dimension of the trapped electron volume perpendicular to the magnetic field (see Figure \ref{fig-zebra-model}). 

The first term in the exponential accounts for the amplitude the power can reach, the second term accounts for the cyclotron modulation via the time dependence of $\Delta'\propto\sin(\ell\omega_{ci}t)$, where the time $t$ is now measured on the scale of the ion-cyclotron period.  Evaluating the exponential term, we find for
\begin{equation}
\mathrm{e}^{2C\tau_\mathit{esc}\Delta'Q_2}=I_0(z)+2\sum\limits_{\eta=1}^\infty I_\eta(z)\exp\left[i\eta\left(\frac{\pi}{2}-\ell\omega_{ci}t\right)\right]
\end{equation}
where $I_\eta(z)$ is the modified Bessel function of order $\eta$, and $z=2\tau_\mathit{esc}CQ_2\beta_E\sqrt{1+R_\mathit{res}^2}$. One observes that the term $I_0(z)$ on the left does not produce any  modulation. Since the power is a real quantity, only the real part of this expression contributes. Thus we obtain
\begin{equation}
{\cal P}(\nu_{ce},\tau_{esc}, t)={\cal P}_0\left\{I_0(z)+\sum\limits_{\eta=1}^\infty I_\eta(z)\sin[\eta(\ell\omega_{ci}t)]\right\} \quad
\end{equation}
where $(\ell,\eta)\in\textsf{N}$, and we have written
\begin{equation}
{\cal P}_0=\frac{1}{\mu_0c}|E_\perp(\nu_{ce})|^2_0\ \exp[2CQ_1\tau_\mathit{esc}]
\end{equation}
These expressions show indeed that the emitted maser power experiences a modulation at multiples of the ion cyclotron frequency. In frequency space $\varpi$, defining  $\xi=\ell\omega_{ci}$, this corresponds to
\begin{eqnarray}
{\cal P}(\nu_{ce},\varpi)&\propto& 2\pi {\cal P}_0\left\{ I_0(z)\delta(\varpi)+\right.\nonumber \\
&+&\left.\sum\limits_{\eta=1}^\infty \left|I_\eta(z)\left[\delta(\varpi-\eta\xi)-\delta(\varpi+\eta\xi)\right]\right|\right\}
\end{eqnarray}
The first term on the right describes just the stationary spectrum while the sum superimposes the ion cyclotron modulation on the spectrum.

The above mechanism, a sketch of which is given for illustration in Figure \ref{fig-zebra-model}, may occasionally work and produce the wanted Zebra structures with periodic amplification and suppression of electron cyclotron maser emission in type IV radio bursts. It requires a rather steep or abrupt transition between the electron distribution and the loss cone. It, however, also depends on the excitation of strong ion cyclotron harmonic waves in order to satisfy the restrictions imposed. This assumes that an intense ion (conic) ring distribution has been generated during the flare, a process that works only in the presence of sufficiently strong field aligned electric fields produced in a magnetic mirror configuration. Because of these reasons it might not be realised very frequently.

\begin{acknowledgements}
This research was part of an occasional Visiting Scientist Programme in 2006/2007 at ISSI, Bern. RT thankfully recognises the assistance of the ISSI librarians, Andrea Fischer and Irmela Schweizer. He highly appreciates the encouragement of Andr\'e Balogh, former Director at ISSI. 
\end{acknowledgements}

\end{document}